\documentstyle[12pt]{article}
\newcommand{\slp}{\raise.15ex\hbox{$/$}\kern-.57em\hbox{$\partial$}}
\newcommand{\sla}{\raise.15ex\hbox{$/$}\kern-.57em\hbox{$a$}}
\newcommand{\slA}{\raise.15ex\hbox{$/$}\kern-.57em\hbox{$A$}}
\newcommand{\slb}{\raise.15ex\hbox{$/$}\kern-.57em\hbox{$b$}}
\newcommand{\be}{\begin{equation}}
\newcommand{\ee}{\end{equation}}
\newcommand{\bear}{\begin{eqnarray}}
\newcommand{\ear}{\end{eqnarray}}
\newcommand{\ba}{\begin{eqnarray*}}
\newcommand{\ea}{\end{eqnarray*}}

\newcommand{\Li}{\cal L}

\begin{document}
%%%%%%%%%%%%%%%%%%%%%%%%%%%%%%%%%%%%%%%%%%%%%%%%%%%%%%%%%%%%%%%%%%%%%%%%%
\begin{titlepage}
\begin{flushright}
HD-THEP--97--61
\\
SOGANG-HEP 227/97
\end{flushright}
\bigskip
\begin{center}
{\bf\LARGE Lagrangian approach to local symmetries and}\\
\vspace{0.5cm}
{\bf \LARGE self-dual model in gauge invariant formulation}\\
\vspace{1cm}
Yong-Wan Kim\footnote{e-mail:~ywkim@physics.sogang.ac.kr,
Department of Physics and Basic Science Research Institute,
Sogang University, C.P.O.Box 1142,
Seoul 100-611, Korea} 
and Klaus D. Rothe\footnote{e-mail:~k.rothe@thphys.uni-heidelberg.de} \\
\bigskip
Institut  f\"ur Theoretische Physik\\
Universit\"at Heidelberg\\
Philosophenweg 16, D-69120 Heidelberg\\
\vspace{1cm}
\end{center}
\begin{abstract}
\noindent Taking the St\"uckelberg Lagrangian associated with the
abelian self-dual model of P. K. Townsend et al as a starting point, 
we embed this mixed first- and second-class system into a pure
first-class system by
following systematically the generalized Hamiltonian approach of  
Batalin, Fradkin and Tyutin. The resulting Lagrangian possesses an
extended gauge invariance and provides a non-trivial example for a general 
Lagrangian approach to unravelling the full set of local symmetries of
a Lagrangian.
\end{abstract}
\vspace{1cm}



Local symmetries; Lagrangian approach;
\end{titlepage}
%%%%%%%%%%%%%%%%%%%%%%%%%%%%%%%%%%%%%%%%%%%%%%%%%%%%%%%%%%%%%%%%%%%%%%%%%%%%%
\setcounter{section}{0}
\section{Introduction}

The Hamiltonian approach to the quantization of constrained systems possessing
first-class constraints has the drawback of not necessarily leading to a
manifestly Lorentz covariant  partition function.  This problem is avoided
in the Lagrangian field-antifield approach \cite{GPS}, 
which is based on an analysis
of the local symmetries of a Lagrangian. The full set of such
symmetries may not be evident from the outset, if the Lagrangian is of a
more complicated nature. Their systematic and exhaustive determination thus
constitute an integral part of the field-antifield quantization program.

In this paper, we reconsider the abelian version of the self-dual (SD)
model of Townsend  et al \cite{TPV}. This model corresponds to a purely 
second-class system in the terminology of Dirac \cite{Di}. Following the
generalized Hamiltonian approach of  Batalin, Fradkin and Tyutin (BFT)
\cite{BFT},
Kim et al \cite{KPKK} have turned all of the second-class constraints into
first-class ones\footnote{For a discussion following the Batalin-Fradkin 
approach see \cite{BRR}}. The resulting Lagrangian was found to be given 
as the sum of a St\"uckelberg type Lagrangian \cite{Goto} and a ``Wess-Zumino"
Lagrangian lacking manifest Lorentz, and even rotational invariance.

On the other hand, the non-abelian SD-model has been shown to be 
equivalently described by
a St\"uckelberg Lagrangian involving mixed first- and second-class 
constraints \cite{KR}. 
Taking the abelian restriction of this Lagrangian as a starting point 
we obtain, following the systematic procedure of BFT, 
an equivalent Lagrangian
involving only first-class constraints, exhibiting manifest rotational
invariance. The equivalence with the formulation of \cite{KPKK}
is established
after suitable integration over one of the auxiliary fields in the extended
phase space. This is done in Sections 2 and 3.

In Section 4, we then take the resulting first-class Lagrangian of Section 3
and systematically unravel all of its local symmetries by following a method
recently discussed in the literature \cite{Shi}. 
Although these symmetries could
of course be derived from the knowledge of the first-class constraints of the
BFT Hamiltonian construction, it turns out that the model provides an
interesting application of the Lagrangian approach of ref. \cite{Shi},
going well beyond the examples discussed in that reference.

We conclude in Section 5 with a summary.

%%%%%%%%%%%%%%%%%%%%%%%%%%%%%%%%%%%%%%%%%%%%%%%%%%%%%%%%%%%%%%%%%%%%%%%%%

\section{First Class Formulation of Self-Dual Model}
\setcounter{equation}{0}
Our starting point is the Lagrangian of the
abelian self-dual model of Townsend et al. \cite{TPV},
\be\label{2.1}
{\Li}_{SD}=\frac{1}{2}m^2B^\mu B_\mu-\frac{m}{2}
\epsilon_{\mu\nu\rho}B^\mu\partial^\nu B^\rho.\ee
This Lagrangian describes a purely second-class system. As was shown
in ref. \cite{KR}, following the procedure of BFT for Hamiltonian
embedding, this Lagrangian turns into a Lagrangian of the St\"uckelberg
type, if one pair of second-class constraints is turned into first-class
constraints. The result of this is a Lagrangian in which the fields
$B^\mu$ have simply been gauged,
\be\label{2.2}
\tilde{\Li}=\frac{1}{2}m^2(B_\mu+\partial_\mu\theta)
(B^\mu+\partial^\mu\theta)-\frac{m}{2}\epsilon_{\mu\nu\rho}
B^\mu\partial^\nu B^\rho.\ee
The Lagrangian $\tilde{\Li}$ describes a mixed system of first- and
second-class constraints. We shall take it as a starting point
for applying the BFT method \cite{BFT} to also convert the
remaining second-class constraints to first-class ones.

From (\ref{2.2}) we obtain for the primary constraints,
\ba
&&T_0:=\pi_0=0,\\
&&T_i:=\pi_i+\frac{1}{2m}\epsilon_{ik}B^k=0.\ea
The canonical Hamiltonian associated with (\ref{2.2}) is given
by
\be\label{2.3}
H_c=\int d^2y\Bigl\{\frac{1}{2}\pi^2_\theta+\frac{1}{2}B_i^2-B^i
\partial_i\theta+\frac{1}{2}(\partial_i\theta)^2
+B^0(-\partial^iT_i+\frac{1}{m}\epsilon_{ij}\partial^iB^j-\pi_\theta
)\Bigr\},\ee
where
\be\label{2.4}
\pi_\theta=\partial^0\theta+B^0.\ee
The Dirac algorithm leads to a secondary constraint
\be\label{2.5}
T_3:=\partial^iT_i-\frac{1}{m}
\epsilon_{ij}\partial^iB^j+\pi_\theta=0.\ee
The constraints $T_0$ and
$T_3$ are seen to be first-class
\bear\label{2.6}
\{T_0(x),\ T_\alpha(y)\}&=&0,\nonumber\\
\{T_3(x),\ T_\alpha(y)\}&=&0,\quad \alpha=0,1,2,3,\ear
whereas the constraints $T_i=0$ are found to be second class.
\be\label{2.7}
\{T_i(x),T_j(y)\}=\frac{1}{m}\epsilon_{ij}\delta^2(x-y)
\equiv \Delta_{ij}(x,y).\ee
The conversion of the second-class constraints $T_i$ to first-class
ones, $\Omega_i$, follows the standard BFT procedure. For the
constraints $T_i$ one introduces ``gauge-degrees of
freedom'' $\phi^i,\ i=1,2$,
together with the symplectic structure
\be\label{2.8}
\{\phi^i(x),\phi^j(y)\}=\omega^{ij}(x,y)\ee
with $\omega^{ij}$ an arbitrary antisymmetric ``matrix'', which we
choose as
\be\label{2.9}
\omega^{ij}(x,y)=-\epsilon^{ij}\delta^2(x-y).\ee
For the first-class constraints $\Omega_i=0$, we make the ansatz
\be\label{2.10}
\Omega_i=\sum^\infty_{n=0}\Omega^{(n)}_i, \
\Omega^{(0)}_i=T_i,\ee
where $\Omega^{(n)}_i$ is a polynomial of degree $n$ in the
new fields $\phi^j$. With the ansatz,
\be\label{2.11}
\Omega_i^{(1)}(x)=\int d^2yX_{ij}(x,y)\phi^j(y),\ee
the requirement of strong involution,
\be\label{2.12}
\{\Omega_i(x),\ \Omega_j(y)\}=0,\ee
then leads to recursive relations, plus the requirement
\be\label{2.13}
\Delta_{ij}(x,y)+\int d^2x'd^2y'X_{ik}(x,x')
\omega^{kl}(x',y')X_{jl}(y',y)=0.\ee
This leads to the solution
\be\label{2.14}
X_{ij}(x,y)=\frac{1}{\sqrt m}\delta_{ij}\delta^2(x-y)\ee
with the inverse
\be\label{2.15}
X^{ij}(x,y)=\sqrt m \delta_{ij}\delta^2(x-y),\ee
as well as to the constraints $\Omega_{\alpha}=0$, with
\bear\label{2.16}
&&\Omega_0=\pi_0,\nonumber\\
&&\Omega_i=\pi_i+\frac{1}{2m}\epsilon_{ij}B^j+
\frac{1}{\sqrt m}\phi^i,\nonumber\\
&&\Omega_3=\partial^i\Omega_i^{(0)}-\frac{1}{m}
\epsilon_{ij}\partial^iB^j+\pi_\theta.\ear
For the first-class Hamiltonian we similarly make the ansatz
\be\label{2.17}
\tilde H=H_c+\sum^\infty_{n=1}H^{(n)}.\ee
The requirement that $\tilde H$ be in strong involution
with the first-class constraints then leads to the recursive relations
\be\label{2.18}
H^{(n)}=-\frac{1}{n}\int d^2xd^2yd^2z\phi^i(x)\omega_{ij}(x,y)
X^{jk}(y,z)G_k^{(n-1)}(z),\quad (n\geq1),\ee
where the generating functionals $G_k^{(n)}$ are given
by
\bear\label{2.19}
&&G_i^{(0)}=\{\Omega_i^{(0)},H_c\}_O,\nonumber\\
&&G_i^{(n)}=\{\Omega_i^{(0)},H^{(n)}\}_O+\{\Omega_i^{(1)},
H^{(n-1)}\}_O,\ear
where the subscript $O$ means that Poisson brackets
are to be calculated
with respect to the original variables. We find recursively
\bear\label{2.20}
&&G_i^{(0)}=B_i+\partial_i\theta,\quad H^{(1)}=\sqrt m\int
d^2x\phi^i\epsilon_{ij}(B^j+\partial^j\theta),\nonumber\\
&&G_i^{(1)}=\sqrt m \epsilon_{ij}\phi^j,\quad H^{(2)}=\frac{m}{2}
\int d^2x \phi^i\phi^i,\nonumber\\
&&G_i^{(n)}=0,\quad H^{(n+1)}=0,\quad n\geq2.\ear
One readily checks that the final Hamiltonian
\be\label{2.21}
\tilde H=H_c+H_1+H_2\ee
is indeed strongly involutive with respect to the constraints.
%%%%%%%%%%%%%%%%%%%%%%%%%%%%%%%%%%%%%%%%%%%%%%%%%%%%%%%%%%%%%%%%%%%%
\section{The Partition Function}
\setcounter{equation}{0}
The partition function corresponding to the Hamiltonian (\ref{2.21})
reads
\be\label{3.1}
Z=\int D\pi_\mu DB^\mu\int D\pi_\theta D
\theta\int D\phi^1D\phi^2
\prod_\alpha\delta[\Omega_\alpha]\prod_\beta\delta[\Gamma_\beta]
det\{\Omega_\alpha,\Gamma_\beta\}e^{iS'},\ee
where
\be\label{3.2}
S'=\int d^3x\{\pi_\mu\dot B^\mu+\pi_\theta\dot\theta+\frac{1}{2}
\phi^i\epsilon_{ij}\dot\phi^j-{\cal H}_c
-\sqrt m \phi^i\epsilon_{ij}(B^j+\partial^j\theta)
-\frac{m}{2}\phi^i\phi^i\},\ee
and $\Gamma_\beta$ are gauge-fixing conditions.
Noting that
\be\label{3.3}
\prod_i\delta[\Omega_i]\delta[\Omega_3]=\prod_i\delta[\Omega_i]\delta
[-\frac{1}{m}\epsilon_{jk}\partial^jB^k+\pi_\rho-
\frac{1}{\sqrt m}\partial^j\phi^k],\ee
and writing the second $\delta$-function as a Fourier integral with
Fourier variable $\xi$, we obtain, after performing the $\pi_\mu$
integration and making the shift of variable $B^0+\xi\to B^0$,
\be\label{3.4}
Z=\int D\xi\int DB^\mu\int D\pi_\theta D\theta
\int\prod^2_{i=1}D\phi^i
\prod_\beta\delta[\Gamma_\beta]det\{\Omega_\alpha,\Gamma_\beta\}
e^{iS''},\ee
where
\be\label{3.5}
S''=S+S_{WZ}\ee
with $S$ the action corresponding to the Lagrangian (\ref{2.2}),
and
\be\label{3.6}
S_{WZ}=\int d^2x\Bigl\{-\frac{1}{\sqrt m}\phi^iB^{0i}-\sqrt m
\phi^i\epsilon_{ij}(B^j+\partial^j\theta)
+\frac{1}{2}\phi^i\epsilon_{ij}\partial_0\phi^j-\frac{m}
{2}\phi^i\phi^i\Bigr\}.\ee
Unlike the action (51) of ref. \cite{KPKK}, the action (\ref{3.5})
is manifestly rotationally invariant. Nevertheless, the corresponding
partition functions are equivalent. Indeed, integration over $\phi^1$
in (\ref{3.4}) turns $Z$ into the partition function of ref. \cite{KPKK}.

To conclude this section, let us show that unlike in the case of the
configuration space partition function of ref. \cite{KPKK}, there
exists a ``canonical'' gauge in which we recover the
partition function of the St\"uckelberg formulation.

In the Hamiltonian formulation, the first-class constraints
(\ref{2.16}) are generators of local symmetry transformations.
Thus, defining
\be\label{3.7}
G=\sum_\alpha\int d^2x\epsilon^\alpha(x)\Omega_\alpha(x),\ee
we have $(\delta A=\{A,g\})$
\bear\label{3.8}
&&\delta B^0=\epsilon^0,\nonumber\\
&&\delta B^i=\epsilon^i+\partial^i\epsilon^3,\nonumber\\
&&\delta\theta=\epsilon^3,\nonumber\\
&&\delta\phi^i=-\frac{1}{\sqrt m}\epsilon_{ij}
\epsilon^j.\ear
One easily checks that the transformation (\ref{3.8})
leaves the action (\ref{3.5}) invariant, provided that we choose
$\epsilon^0=\partial_0\epsilon^3$\footnote{As is well
known, symmetry transformations of the Lagrangian generally imply
restrictions on the symmetry transformations of the total Hamiltonian.}.
Consider in particular the transformation with
$\epsilon^0=\epsilon^3=0$, and
$\epsilon^j=\sqrt m\epsilon_{jk}\phi^k$. Under this transformation
$\phi^i\to0,\ i=1,2$; the gauge
$\phi^1=\phi^2=0$ is thus accessible, and the partition function
(\ref{3.4}) reduces in this gauge to that of the St\"uckelberg
formulation (\ref{2.2}).
%%%%%%%%%%%%%%%%%%%%%%%%%%%%%%%%%%%%%%%%%%%%%%%%%%%%%%%%%%%%%%%%%%%%%%%%
\section{Lagrangian Approach to Local Symmetries}
\setcounter{equation}{0}

In the field-antifield formalism \cite{GPS} the establishment
of the full, irreducible set of local symmetries of a Lagrangian
plays a fundamental role. In this formalism these symmetries
must be identified without the use of
Dirac's Hamiltonian construction of the corresponding generators.
In this section we illustrate a recently proposed \cite{Shi}
Lagrangian approach in terms of the action (\ref{3.5}) in
the extended configuration space. Our notation deviates
somewhat from that of ref. \cite{Shi}.

The equations of motion following from (\ref{3.5}) are
of the form
\be\label{4.1}
L_{ix}=\int d^3y (W_{ix,jy}\ddot\varphi_{jy}+\alpha_{ix})=0, \
i=1,2,...,6,\ee
where
\be\label{4.2}
(\varphi_{ix})^{\rm T}
=(B^0(x),B^1(x),B^2(x),\phi^1(x),\phi^2(x),\theta(x)),\ee
$W_{ix,jy}$ is the Hessian
\bear\label{4.3}
W_{ix,jy}&=&\frac{\delta^2L}{\delta\dot\varphi_{ix}\delta\dot
\varphi_{jy}}=\delta_{i6}\delta_{j6}\delta^3(x-y)\nonumber\\
&\equiv&\tilde W_{ij}\delta^3(x-y),\ear
and
\be\label{4.4}
(\alpha_{ix})^{\rm T}=(\alpha_{B^0}(x),\alpha_{B^1}(x),\alpha_{B^2}(x),\alpha
_{\phi^1}(x),\alpha_{\phi^2}(x),\alpha_\theta(x))\ee
with
\bear\label{4.5}
&&\alpha_{B^0}=\frac{1}{m}\epsilon_{ij}\partial^iB^j-(B^0+\partial^0\theta)
+\frac{1}{\sqrt m}\partial^i\phi^i,\nonumber\\
&&\alpha_{B^i}=-\frac{1}{m}\epsilon_{ij}(\partial^0B^j-\partial^jB^0)
+(B^i+\partial^i\theta)-\frac{1}{\sqrt m}
\partial^0\phi^i-\sqrt m\epsilon
_{ij}\phi^j,\nonumber\\
&&\alpha_{\phi^i}=-\epsilon_{ij}\partial_0\phi^j+\frac{1}{\sqrt m}
(\partial^0B^i-\partial^iB^0)+\sqrt m\epsilon_{ij}(B^j+\partial^j
\theta)+m\phi^i,\nonumber\\
&&\alpha_\theta=\partial^0B^0+\partial_i(B^i+\partial^i\theta)
+\sqrt m\epsilon_{ij}\partial^i\phi^j.\ear
The Hessian matrix $\tilde W_{ij}$ (\ref{4.3}) is of rank one.
Hence there exist five ``zeroth generation'' null eigenvectors
$(\lambda_{ix})$ satisfying
\be\label{4.6}
\int d^3x\ \lambda_{ix}\ W_{ix,jy}=0.\ee
We choose them to have components
\be\label{4.7}
\lambda_{ix}(a,z)=\delta^3(x-z)\delta_{ia},\quad a=1,...,5,\ee
where $(a,z)$ label the eigenvectors and $(i,x)$ label
the components.

Correspondingly we have the ``zeroth generation'' constraints
\be\label{4.8}
\Omega_a(z)=\int d^3x\lambda_{ix}(a,z)L_{ix}=\alpha_a(z)=0,\quad
a=1,...,5.\ee
From here on we shall drop continuum labels, which are now implicitly
contained in the associated discrete labels. From (\ref{4.4})
we see that only three of these represent non-trivial constraints, since
we have identically
\be\label{4.9}
\alpha_{B^i}+\frac{1}{\sqrt m}\epsilon_{ij}\alpha_{\phi^j}\equiv0.\ee
Introducing the null eigenvectors
\bear\label{4.10}
&&\hat\lambda(1)=(0,1,0,0,\frac{1}{\sqrt m},0),\nonumber\\
&&\hat\lambda(2)=(0,0,1,\frac{-1}{\sqrt m},0,0),\ear
we can write the identities in the form
\be\label{4.11}
\hat\Omega_{\hat a}:=\hat\lambda_i(\hat a)L_i\equiv0,\ \hat a=1,2.\ee
For the remaining, non-trivial constraints we choose
\be\label{4.12}
\bar\Omega_{\bar a}:=\bar\lambda_i(\bar a)L_i=0,\ \bar
a=1,2,3\ee
with
\be\label{4.13}
\bar\lambda_i(\bar a)=\lambda_i(a).\ee
This also establishes our notation: quantities related to
trivial (non-trivial) constraints are denoted by a ``hat'' (``bar'').

We now require the (non-trivial) constraints (\ref{4.12})
to be independent of time. We thus need to add to the equations
of motion (\ref{4.1}) the equations $\dot{\bar\Omega}_{\bar a}=0,
\ \bar a=1,2,3$. The resulting set of 9 equations may be summarized
in the form of the set of ``first generation'' equations
$L_{i_1}^{(1)}=0,\ i_1=1,...,9$, with
\be\label{4.14}
L_{i_1}^{(1)}=\left\{\begin{array}{ll}
L_i,\ i=1,...6,\\
\frac{d}{dt}(\bar\lambda_i(\bar a)L_i),\ i=6+\bar a,\ \bar a=1,2,3.
\end{array}\right..\ee
$L_{i_1}^{(1)}$  can be written in the general form
\be\label{4.15}
L_{i_1}^{(1)}:=W_{i_1j}^{(1)}\ddot\varphi_j+\alpha^{(1)}_{i_1}=0,\ee
where
\be\label{4.16}
(W^{(1)}_{i_1j})=\left(\begin{array}{cc}
\\
\\
(\tilde W_{ij})\\
\\
\hline\\
\begin{array}{cccccc}
0&0&0&0&0&1\\
0&0&\frac{1}{m}&\frac{1}{\sqrt m}&0&0\\
0&-\frac{1}{m}&0&0&\frac{1}{\sqrt m}&0\end{array}\end{array}
\right)\delta^2(x-y),\ee
and
\be\label{4.17}
(\alpha^{(1)}_{i_1})=\left(\begin{array}{c}
\\
\\
(\alpha_i)\\
\\
\\
\hline\\
\alpha_7^{(1)}\\
\alpha_8^{(1)}\\
\alpha_9^{(1)}\end{array}\right)\ee
with
\bear\label{4.18}
&&\alpha^{(1)}_7=-\frac{1}{m}\partial^1\dot B^2+\frac{1}{m}\partial^2
\dot B^1+\dot B^0-\frac{1}{\sqrt m}\partial^i\dot\phi^i,\nonumber\\
&&\alpha^{(1)}_8=-\frac{1}{m}\partial^2\dot B^0-
(\dot B^1+\partial^1\dot\theta)+\sqrt m\dot\phi^2,\nonumber\\
&&\alpha^{(1)}_9=\frac{1}{m}\partial^1\dot B^0-(
\dot B^2+\partial^2\dot\theta)-{\sqrt m}\dot\phi^1,\ear
respectively.

We now repeat the previous analysis taking eqns. (\ref{4.15}) as a
starting point, and looking for solutions of
\be\label{4.19}
\lambda^{(1)}_{i_1}(a_1)W^{(1)}_{i_1j}=0.\ee
The null eigenvectors of $W^{(1)}$ are of the
form $(\lambda_1,\lambda_2,\lambda_3,\lambda_4,\lambda_5,
\lambda_6,-\lambda_4,0,0)$.
In addition to $(\vec\lambda(\bar a),0,0,0),\ \bar a=1,2,3$
and $(\vec\lambda(\hat a),0,0,0),\ \hat a=1,2,$
we have one further null eigenvector, which we choose as
\footnote{Each component is understood to be multiplied
by $\delta^2(x-z)$.}
\be\label{4.20}
\lambda^{(1)}(1)=(0,0,0,0,0,1,-1,0,0).\ee
Associated with this eigenvector we have a new constraint
\be\label{4.21}
\Omega_1^{(1)}:=\lambda^{(1)}_{i_1}\alpha^{(1)}_{i_1}=0.\ee
Explicitly
\bear\label{4.22}
\Omega_1^{(1)}&=&\alpha^{(1)}_6-\alpha_7^{(1)}\\
&=&\partial_i(B^i+\partial^i\theta)+\sqrt m\epsilon_{ij}\partial^i\phi^j+
\frac{1}{m}\epsilon_{ij}\partial^i\dot B_j+\frac{1}{\sqrt m}
\partial^i\dot\phi^i.\nonumber\ear
Since this represents a non-trivial constraint, we write $\Omega_1
^{(1)}=\bar\Omega_{\bar 1}
^{(1)},\ \lambda^{(1)}(1)=\bar\lambda^{(1)}(\bar 1)$.
Correspondingly (\ref{4.21}) may be written in the form
\be\label{4.23}
\bar\Omega^{(1)}_{\bar a_1}=\bar\lambda^{(1)}_{i_1}
(\bar a_1)L^{(1)}_{i_1}=0,
\quad \bar a_1=1.\ee
There exist no new identities at this level.

We repeat the procedure by requiring $\dot{\bar\Omega}^{(1)}_{\bar a_1}
=0$, i.e.
\be\label{4.24}
\frac{1}{\sqrt m}\partial^i\ddot\phi^i+\frac{1}{m}\epsilon
_{ij}\partial^i\ddot B^j+\partial_i(\dot B^i+
\partial^i\dot\theta)+\sqrt m\epsilon_{ij}\partial^i\dot\phi^j=0.\ee
Adjoining this equation to $L_{i_1}^{(1)}=0$,
we have $L^{(2)}_{i_2}=0,\ i_2=1,...,10,$ with
\be\label{4.25}
L^{(2)}_{i_2}=\left\{\begin{array}{lll}
L_i,\ i=1,...,6,\\
\frac{d}{dt}(\bar\lambda_i(\bar a)L_i),\ i_2=6+\bar a, \ \bar a=1,2,3,\\
\frac{d}{dt}(\bar\lambda_{i_1}^{(1)}(\bar a_1)L^{(1)}_{i_1}),\
i_2=9+\bar a_1, \ \bar a_1=1.\end{array}\right.\ee
The resulting complete set of equations is now of the form
\be\label{4.26}
L^{(2)}_{i_2}:=W^{(2)}_{i_2j}\ddot\varphi_j+\alpha^{(2)}_{i_2}=0,
\quad i_2=1,...,10,\ee
where
\bear\label{4.27}
&&W^{(2)}_{i_2j}=W_{i_2j},\quad i_2=1,...,6,\nonumber\\
&&W^{(2)}_{i_2j}=W_{i_2j}^{(1)},\quad i_2=7,8,9,\\
&&W^{(2)}_{10,j}=(0,-\frac{\partial^2}{m},
\frac{\partial^1}{m},\frac{\partial
^1}{\sqrt m},\frac{\partial^2}{\sqrt m},0)\delta^2(x-y),\nonumber\ear
and
\bear\label{4.28}
&&\alpha^{(2)}_{i_2}=\alpha_{i_2},\quad i_2=1,...,6,\nonumber\\
&&\alpha^{(2)}_{i_2}=\alpha_{i_2}^{(1)},\quad i_2=7,8,9,\\
&&\alpha^{(2)}_{10}=\partial_i(\dot B^i+\partial^i
\dot\theta)+\sqrt m\epsilon_{ij}\partial^i\dot\phi^j.\nonumber\ear
In addition to the previous null
eigenvectors (enhanced by a suitable number of zeroes in the last
components), we thus have the new null eigenvector
\be\label{4.29}
\lambda^{(2)}(1)=(0,0,0,0,0,0,0,\partial^1_x,\partial^2_x,-1)
\delta^2(x-z).\ee
The associated constraint is found to vanish ``identically'':
\be\label{4.30}
\lambda^{(2)}_{i_2}(1)\alpha^{(2)}_{i_2}=\partial^1\alpha^{(2)}_8
+\partial^2\alpha^{(2)}_9-\alpha_{10}^{(2)}=0.\ee
Hence we denote $\lambda^{(2)}(1)$ by $\hat\lambda^{(2)}(\hat 1)$
and write (\ref{4.30}) in the form
\be\label{4.31}
\hat\Omega^{(2)}_{\hat a_2}:=\hat\lambda^{(2)}_{i_2}(\hat a_2)L^{(2)}_{i_2}
\equiv0,\quad \hat a_2=1.\ee
The algorithm ends at this point.

The local symmetries of the action (\ref{3.5}) are encoded in the identities
(\ref{4.11}) and (\ref{4.31}). Recalling (\ref{4.14}) and (\ref{4.25})
we see that these identities can be rewritten as follows:
\bear\label{4.32}
&&\hat\Omega_1=L_2+\frac{1}{\sqrt m}L_5\equiv0,\nonumber\\
&&\hat\Omega_2=L_3-\frac{1}{\sqrt m}L_4\equiv0,\\
&&\hat\Omega^{(2)}_1=\frac{d}{dt}(\partial^1L_2+\partial^2L_3+L_6)+\frac
{d^2}{dt^2}L_1\equiv0.\nonumber\ear
This result is a special case of a general theorem stating
\cite{Shi} that the identities $\hat\Omega^{(l)}_{\hat a_l}\equiv
0 \ (\hat\Omega^{(0)}_{\hat a_0}\equiv\hat\Omega_{\hat a})$ can always be
written in the form
\be\label{4.33}
\hat\Omega^{(l)}_{\hat a_l}=\sum^l_{s=0}
\frac{d^s}{dt^s}\left(\Phi^{(l)}_{si}(\hat a_l)L_i\right).\ee
For the case in question
\bear\label{4.34}
&&\Phi^{(0)}_{0i}(1)=\delta_{i2}+\frac{1}{\sqrt m}\delta_{i5},\nonumber\\
&&\Phi^{(0)}_{0i}(2)=\delta_{i3}-\frac{1}{\sqrt m}\delta_{i4},\nonumber\\
&&\Phi^{(1)}_{1i}(1)=\delta_{i2}\partial^1+\delta_{i3}
\partial^2+\delta_{i6},\nonumber\\
&&\Phi^{(2)}_{0i}(1)=0\nonumber\\
&&\Phi^{(2)}_{2i}(1)=\delta_{i1}.\ear
It again follows from general considerations \cite{Shi}
that the action (\ref{3.5}) is invariant under the transformation
\be\label{4.35}
\delta\varphi_i=\sum_{l=0}\sum^l_{s=0}(\sum_{\hat a_l}(-1)^s\Phi_{si}
^{(l)}(\hat a_l)\frac{d^s}{dt^s}\epsilon^{(l)}_{\hat a_l}).\ee
For the case in question this corresponds to the transformations
\bear\label{4.36}
&&\delta B^0=\ddot\epsilon^{(2)}_1,\nonumber\\
&&\delta B^1=\epsilon^{(0)}_1-\partial^1\dot\epsilon^{(1)}_1,\nonumber\\
&&\delta B^2=\epsilon_2^{(0)}-\partial^2\dot\epsilon^{(1)}_1,\nonumber\\
&&\delta\phi^1=-\frac{1}{\sqrt m}\epsilon_2^{(0)},\nonumber\\
&&\delta\phi^2=\frac{1}{\sqrt m}\epsilon^{(0)}_1,\nonumber\\
&&\delta\theta=-\dot\epsilon^{(1)}_1.\ear
Comparison with (\ref{3.8}) with $\epsilon^0=\ddot\epsilon_1^{(2)},
\epsilon^3=-\dot\epsilon^{(1)}_1$, and $\epsilon^i
=\epsilon^{(0)}_i$ shows that we have recovered
the local symmetry transformations.
%%%%%%%%%%%%%%%%%%%%%%%%%%%%%%%%%%%%%%%%%%%%%%%%%%%%%%%%%%%%%%%%%%%%%%%%%%%%%

\section{Conclusion}
In this paper we reconsidered the abelian self-dual model of Townsend et al
\cite{TPV}, by taking as a starting point its description as a mixed 
first- and second-class system in terms of a St\"uckelberg type 
Lagrangian \cite{Goto}.
The equivalence of this description with the original, 
purely second-class system
was established in \cite{KR} for the non-abelian case, 
and hence in particular
for the abelian case. We turned this St\"uckelberg Lagrangian into a purely
first-class system, following the method of Batalin, Fradkin and
Tyutin \cite{BFT}. 
We then showed that the Lagrangian thus obtained was equivalent
to the Lagrangian eq. (51) of ref. \cite{KPKK}. 
This established that despite the lack of manifest rotational 
invariance of that Lagrangian, 
it defined a theory consistent with rotational invariance.

The Lagrangian we obtained in this paper 
exhibited a larger local symmetry than the original
St\"uckelberg Lagrangian, reflecting the existence of twice as many
first-class constraints, that is, generators of gauge transformations.
We then showed, following the method of ref. \cite{Shi}, how these symmetries could be systematically derived on a
purely Lagrangian level, without resorting to a Hamiltonian formulation.
The Lagrangian approach employed in this derivation should be of much interest in the context of the field-antifield formalism.
As it turned out, our formulation of the self-dual model 
as a purely first-class system provided a non-trivial example 
for such a systematic construction.

%%%%%%%%%%%%%%%%%%%%%%%%%%%%%%%%%%%%%%%%%%%%%%%%%%%%%%%%%%%%%%%%%%%%%%%%%%%%%

\section*{Acknowledgement}
One of the authors (Y. W. Kim) would
like to thank the Institut f\"ur Theoretische Physik for the
kind hospitality, and the Korea Research Foundation for  
(1996) overseas fellowship 
which made this collaboration possible.

%%%%%%%%%%%%%%%%%%%%%%%%%%%%%%%%%%%%%%%%%%%%%%%%%%%%%%%%%%%%%%%%%%%%%%%%%%%%%

\end{document}